\documentclass[a4paper,10pt]{article}

\def\eq#1{{Eq.~(\ref{#1})}}

\newcommand{\LL}{Lanczos-Lovelock}

\newcommand{\Cal}[1]{\ensuremath{\mathcal{#1}}}
\newcommand{\ph}[1]{\phantom{#1}}
\newcommand{\D}{\ensuremath{\nabla}}

\newcommand{\cc}{cosmological constant}

\title{Equipartition of energy in the horizon degrees of freedom and the emergence of gravity}
\author{T. Padmanabhan\\IUCAA, Pune University Campus,\\
Ganeshkhind, Pune 411007, INDIA\\
email: nabhan@iucaa.ernet.in}
\date{ }

\begin{document}

\maketitle

\begin{abstract}
It is possible to provide a physical interpretation for the field equations of gravity
based on a thermodynamical perspective. The virtual degrees of freedom associated with the horizons, as perceived by the local  Rindler observer, play a crucial role in this approach. In this context, the relation $S=E/2T$ between the entropy ($S$), active gravitational mass ($E$) and temperature ($T$) --- obtained previously in gr-qc/0308070 [\textit{CQG}, \textbf{21}, 4485 (2004)] --- can be reinterpreted as the 
law of equipartition $E = (1/2) nk_BT$  where $n=\Delta A/L_P^2$ is the number (density) of microscopic horizon degrees of freedom in an area $\Delta A$. Conversely, one can use the equipartition argument to provide a thermodynamic interpretation of gravity, even in the non-relativistic limit. These results emphasize the intrinsic quantum nature of all gravitational phenomena and diminishes the distinction between thermal phenomena 
associated with local Rindler horizons and the usual thermodynamics of macroscopic bodies in non-inertial frames.
Just like the original thermodynamic interpretation, these results also hold for  a wide class of gravitational theories like the \LL\ models. 
\end{abstract}

\section{Introduction}

Recent investigations have shown that gravitational field equations in a wide class of models (including Einstein's gravity) can be given a physical interpretation which is 
thermodynamical in origin. In this  approach, one introduces around any event in spacetime local Rindler observers who perceive an acceleration horizon with a temperature $T = \kappa/2\pi$ where $\kappa $ is the acceleration. (We use units with $\hbar=c=k_B=1$  and $ G=L_P^2$ except when stated otherwise. The signature is -- + + +; Greek indices cover space indices and Latin indices cover spacetime indices). The validity of standard laws of thermodynamics in the local Rindler frame can then be related to the field equations of gravity in any diffeomorphism invariant theory. The details of this approach have been presented in several papers \cite{bg} and in the recent reviews \cite{reviews1,reviews2}.
The following features  play a  key role in this  alternative perspective on gravitational interaction.

\begin{enumerate}

\item 
The existence of local Rindler horizons brings in a new level of observer dependence into the theory. For example, when an amount of energy $\delta E$ reaches the local Rindler horizon with temperature $T$, the resultant entropy change,
 as perceived by the Rindler observer,
 will be $\delta S = \delta E/T$. 
The Rindler  observer will interpret this transfer of energy to the horizon degrees of freedom as taking place when  the matter is within a few Planck lengths from the horizon, though 
it takes (formally) infinite amount of coordinate time for the matter to actually reach the horizon.
The inertial observer
 will not share this point of view, thereby introducing the new level of observer dependence
 in the thermodynamic description of even `normal' systems when viewed in an non-inertial frame.

\item
One can relate the thermodynamical parameters (like entropy $S$ and temperature $T$) associated with the horizon to the active gravitational mass $E$ producing the gravitational acceleration in the spacetime. In the case of static spacetimes in general relativity, this relation
turns out to be 
\begin{equation}
S=\frac{E}{2T} \equiv \frac{1}{2} \beta E
\label{one}
\end{equation} 
This was obtained and discussed in detail in Ref. \cite{cqgpap}. This result continues to hold for a much wider class of theories like the \LL\ models of gravity (see Section 5.8 of Ref. \cite{reviews1}).

\item
Because gravity is an emergent phenomenon, the field equations governing gravity can be obtained from an entropy maximization principle without varying the metric tensor \cite{entropy}. Exremising the expression for spacetime entropy
\begin{equation}
S[u^a]\propto\int_\Cal{V}{d^Dx\sqrt{-g}}
    \left(4P_{ab}^{\ph{a}\ph{b}cd} \D_cu^a\D_du^b \right) \,,
\label{ent-func-2}
\end{equation}
with respect to $u^a$
subject to the constraints that (i) the integral over  $T_{ab}u^au^b$, is a constant and (ii) $\delta(u_au^a)=0$  will lead to the field equations $\mathcal{G}_{ab} = 8\pi L_P^2 (T_{ab}+\Lambda g_{ab})$ for \LL\ gravity with a \cc\ $\Lambda$ 
where 
\begin{equation}
\mathcal{G}_{ab}
\equiv P_a^{\phantom{a} cde} R_{bcde}  - \frac{1}{2} L g_{ab}
\equiv\mathcal{R}_{ab}-\frac{1}{2} L g_{ab};\quad
P^{abcd} \equiv \frac{\partial L}{\partial R_{abcd}}
\label{genEab1}
\end{equation}
and $L$ is the \LL\ Lagrangian. (This will reduce to Einstein's theory in $D=4$). 
This variational principle can be thought of as extremising the \textit{total} entropy of matter and gravity when $u_a$ is a null vector. But it can also be interpreted (along the lines of ref.\cite{cqgpap}), in static spacetimes with a horizon, as exremising the \textit{gravitational} entropy in \eq{ent-func-2} subject to the constraints that (i) $\delta(u_iu^i)=0$   and (ii) the total matter energy $U$ is constant. Such spacetimes are described by the line element
\begin{equation}
ds^2=-N^2(\mathbf{x})dt^2+\gamma_{\mu\nu}(\mathbf{x}) dx^\mu dx^\nu
\label{dssquare}
\end{equation} 
If  $u^i=\xi^i/N$ denotes the four velocity of static observers, where $\xi^i$ is a timelike Killing vector such that $\xi^i\xi_i\equiv N^2(\mathbf{x})=0$ is the location of the horizon $\mathcal{H}$, then the matter energy is given by the integral of $dU=T_{ab}\xi^au^b\sqrt{\gamma}d^{D-1}x
=T_{ab}u^au^b\sqrt{-g}d^{D-1}x$. While
extremising $S-\beta U$, we can identify $\beta$ with the range of time integration by analytic continuation from the Euclidean sector so that $\beta U$ becomes an integral over
$T_{ab}u^au^b\sqrt{-g}d^Dx$.
Further, as explained in section 7.2 of ref. \cite{reviews1}, one can use the identity
\begin{equation}
\label{details1}
4P_{ab}^{\ph{a}\ph{b}cd} \D_cu^a\D_du^b=4\D_c[P_{ab}^{\ph{a}\ph{b}cd} u^a\D_du^b]+2\mathcal{R}_{ai}u^au^i
\end{equation}
to work with an alternative definition of $S$ given by
\begin{equation}
S[u^a]\propto\int_\Cal{V}{d^Dx\sqrt{-g}}
    \left( 2\mathcal{R}_{ai}u^au^i\right) 
\label{altS}
\end{equation}
which, in the context of Einstein's theory (with $\mathcal{R}_{ai}=R_{ai}$) reduces to the expression used in ref.\cite{cqgpap}.

\item
In this approach, the entropy of a region of spacetime resides in its boundary when the field equations are satisfied, making the theory intrinsically holographic. What is more, the surface term in the action, that leads to the entropy, appears as the phase of the semi-classical wavefunction in a quantum theory of gravity \cite{entropyquant}. Applying the Bohr-Sommerfeld quantization condition to this phase shows that the changes in the dimensionless gravitational entropy satisfy the quantization condition 
$\delta S = 2\pi$. In the case of Einstein's theory, entropy (and the relevant surface term) are proportional to the area and this result can be interpreted as a semi-classical area quantization. But a study of  \LL\ models  shows that, in the general context, it is the entropy which is quantized rather than the area \cite{entropyquant}.

\end{enumerate}
The purpose of this short note is to point out that the above results --- especially \eq{one} --- lead to an interesting physical picture as regards the equipartition of energy among the microscopic horizon degrees of freedom  perceived by the local Rindler observer and stress the connection between the field equations and equipartition of energy. 

\section{Equipartition of energy in the horizon degrees of freedom}

We have considerable evidence of very different nature \cite{tplimitations} to suggest that, in a large class of models, the Planck length acts as lower bound to the length scales that can be operationally defined. Suppose we have {\it any} formalism of quantum gravity in
which there is a minimum quantum of length or area, of the order of $L_P^2\equiv G\hbar/c^3$. 
Given this result, 
a patch of horizon having an
area $A$
can be divided into $n=({A}/c_1L_P^2)$ microscopic cells where $c_1$ is a numerical factor. If each cell has $c_2$ internal states
 then the total number of microscopic states is $c_2^n$ and the resulting entropy is
$S=n\ln c_2=(4\ln c_2/c_1)(A/4L_P^2)$. This will lead to the standard result in Einstein's theory, $S=(A/4L_P^2)$ if we choose $(4\ln c_2/c_1)=1$. 

In the thermodynamical approach, fluctuations of the area elements in the hot Rindler horizon play a crucial role. Taking this analogy further, we will assume that the each cell of area  $c_1L_P^2$ contributes an energy $(1/2)T$ in accordance with the standard thermodynamic equipartition law. Then the total energy is 
\begin{equation}
\mathcal{E}\equiv \frac{1}{2}nT=\frac{1}{2}\frac{ST}{\ln c_2}=\frac{2ST}{c_1}=\frac{E}{c_1}
\end{equation} 
We have used $c_1=4\ln c_2$ in arriving at the third equality and \eq{one} to obtain the last equality. This result shows that, if we take the patches to be of size $L_P^2$ (that is, $c_1=1$), then \textit{the equipartition energy of the horizon maches with the active gravitational mass producing the horizon!} Even for other choices of $c_1$ we get the result that if we attribute an energy $(1/2)T$ to each patch of area $L_P^2$, then the equipartition energy matches with the active mass.

One amusing  application of this equipartition theorem is the derivation the gravitational acceleration $\kappa$ produced by a massive spherical body of mass $M$. Though this is conceptually trivial --- because we know (from e.g., \cite{reviews1,reviews2}) that fully relativistic gravity itself has a thermodynamic derivation --- it illustrates the role of the local Rindler temperature in  the thermodynamic interpretation  in a transparent manner. Consider a spherical surface of area $A$ around the massive body. It was argued in Ref. \cite{cqgpap} (see e.g. item(a) on page 4488) that one can also associate with this area the degrees of freedom corresponding to an entropy $S=A/4L_P^2$ because observers at rest on this surface will experience an acceleration produced by the gravitating body. (In fact, one can always construct local Rindler observers who will perceive any sufficiently small patch of a timelike surface as a stretched horizon.) Hence the equipartition energy for the degrees of freedom associated with this surface will give (taking $c_1=1$ or, equivalently, attributing $(1/2)k_BT$
per Planck area), in normal units,
\begin{equation}
\mathcal{E}=\frac{1}{2}\frac{A}{L_P^2}k_BT=\frac{1}{2}\frac{A}{L_P^2}\frac{\hbar \kappa}{2\pi c}
=\frac{A}{4\pi}\frac{c^2\kappa}{G}
\end{equation} 
where we have used the expression of the Rindler temperature $k_BT=\hbar \kappa/2\pi c$ corresponding to the acceleration $\kappa$ of a body at rest on a surface of area $A$. 
Note that $\mathcal{E}$ is independent of $\hbar$ \textit{if} we treat, as is usually done, $G$ and $\hbar$ as unrelated \textit{independent} constants.  The effective number of degrees of freedom on the surface $n\propto 1/\hbar$ while the Rindler temperature $T\propto \hbar$ making the equipartition energy $(1/2)nk_BT$ independent of $\hbar$. Equating the equipartition energy 
$\mathcal{E}$ to the energy of active gravitational mass $Mc^2$, we find that the acceleration induced on a body at rest on a surface of area $A$ is given by
\begin{equation}
\kappa = GM \left( \frac{4\pi}{A}\right) = \frac{GM}{r^2} = \left( \frac{\mathcal{A}_Pc^3}{\hbar}\right)\,  \frac{M}{r^2}
\label{acc}
\end{equation} 
which is just the Newton's law of gravity. 
(The interpretation of this result is identical to the ones provided in Ref.\cite{cqgpap}. 
In fact, \eq{acc} itself is derived, in a related context, in page 4492 of Ref.\cite{cqgpap}.)

The last equality in \eq{acc} shows that, in this holographic approach to gravity, the Planck area $\mathcal{A}_P = L_P^2$ is what determines the gravitational force. This has important implications as regards the classical limit obtained by $\hbar \to 0$. If we keep $\mathcal{A}_P$ finite and take the  limit $\hbar \to 0$ then the coupling constant diverges in the Newton's law, showing that gravity is intrinsically a quantum phenomenon.
(This is similar to the fact that, if we take the strict  $\hbar \to 0$ in any solid state phenomenon, we will obtain a divergent result because the individual atoms making up the solid cannot exist as stable structures in the $\hbar \to 0$ limit.) This point of view and its implications have been discussed earlier in Ref. \cite{intgrav}.

The law of equipartition leads to the field equations of gravity even in the more general context of a static spacetime with a horizon with a line element of the form
in \eq{dssquare}.
 The comoving observers at $x^\mu=$ constant have the four velocity
$u_i=-N\delta^0_i$ and the four acceleration $a^i=(0,\partial^\mu N/N)$. If $N\to 0$ on a surface $\mathcal{H}$, with
$ N^2 a^2\equiv(\gamma_{\mu\nu}\partial^\mu N\partial^\nu N)\to \kappa^2$ finite, then this coordinate system has a horizon and 
one can associate a temperature $|\kappa|/2\pi$ with this horizon. The \textit{local} redshifted inverse temperature is given by $\beta_{loc}\equiv N\beta=2\pi N/|\kappa|$.
The entropy associated with a two-surface, $\partial \mathcal{V}$, bounding a three-volume
$\mathcal{V}$ is defined (see eq. (2) of ref. \cite{cqgpap} which is the same as \eq{altS} in the context of Einstein's theory)  as
\begin{eqnarray}
S &=&  \frac{1}{8\pi G} \int R_{ab} u^a u^b \sqrt{-g} \, d^4 x
= \frac{1}{8\pi G}\int\sqrt{-g}d^4x\nabla_i a^i\nonumber\\
&=&\frac{\beta}{8\pi G}\int_{\partial\cal V}\sqrt{\sigma}d^2x(Nn_\mu a^\mu)
\label{defs}
\end{eqnarray}
while the active gravitational mass-energy is defined by (see eq. (8) of ref. \cite{cqgpap})
\begin{equation}
\label{defe}
E=2\int_{\cal V}d^3x\sqrt{\gamma}N (T_{ab}-\frac{1}{2}Tg_{ab})u^au^b
\equiv 2\int_{\cal V}d^3x\sqrt{\gamma}N (\bar T_{ab}u^au^b)
\end{equation}
The validity of the equipartition condition in \eq{one}, for arbitrary three-volumes $\mathcal{V}$  leads to the field equation $\nabla_i a^i\equiv R_{ab}u^au^b=8\pi G \bar T_{ab}u^au^b$ for the static observers which is essentially the $00$ component of the Einstein equation. Demanding local Lorentz invariance leads to $R_{ab}=8\pi G \bar T_{ab}$.
(In the nonrelativistic limit,  $R_{00}=8\pi G \bar T_{00}$ will lead to the standard Poisson equation. Thus one can obtain the standard equations of gravity from the equipartition argument without assuming spherical symmetry etc.)

In the general case, the equipartition law $S=\beta E/2$ of ref.\cite{cqgpap} itself takes the interesting form of an integral 
over the local acceleration temperature $T_{\rm loc}\equiv (N a^\mu n_\mu) /2\pi$ and is 
given by 
\begin{equation}
E  =  \frac{1}{2}k_B
 \int_{\partial\cal V}\frac{\sqrt{\sigma}\, d^2x}{L_P^2}\left\{\frac{N a^\mu n_\mu}{2\pi}
\right\}
\equiv \frac{1}{2} k_B \int_{\partial\cal V}dn\, T_{\rm loc}
\label{idn}
\end{equation} 
thereby identifying the number of degrees of freedom to be $dn=  \sqrt{\sigma}\, d^2x/L_P^2$ in an area element $\sqrt{\sigma} \, d^2 x$. (One can, alternatively, write down this relation `by inspection' and obtain the gravitational field equations as a consequence, provided one accepts the choice of various numerical factors.)
On the horizon, the factor within curly brackets in \eq{idn} is constant and we recover $E=(1/2)k_B T (A/L_P^2)=2ST$. On any other surface, we get 
the sum of $\Delta E= (1/2) k_B T_{\rm loc} \Delta n$ which is physically reasonable and represents
 the equipartition of microscopic degrees of freedom with the local Rindler temperature.

Since the thermodynamic interpretation of gravitational field equations holds for a wider class of theories like \LL\ models, we would expect the equipartition argument also to generalize for these models.  The procedure to obtain such a generalization using Noether current is given explicitly in Section 5.8 of Ref. \cite{reviews1}. In particular, the relation $S = E/2T$ holds in all these models when we use \eq{altS}. 
But the connection between equipartition energy and the horizon area will not hold in the \LL\ models because the expression for entropy is no longer proportional to the area. This is to be expected since the classical gravitational law will be modified in the \LL\ models. The detailed investigation of these aspects as well as the study of non-static spacetimes will be presented elsewhere.

\section{Conclusions}

The existence of virtual degrees of freedom on the local Rindler horizon and the holographic relationship they hold with respect to the bulk degrees of freedom provide new insights into the nature of gravity. I have emphasized in the previous papers (see e.g., \cite{reviews1,reviews2}) that the behaviour of bulk spacetime is similar to the behaviour of a macroscopic body of, say, gas and can be usefully described through thermodynamic concepts --- even though these concepts may not have any meaning in terms of the true microscopic degrees of freedom. This is exactly similar to the fact that, while one cannot  attribute entropy, pressure or temperature to a single molecule of gas, they are useful quantities to describe the bulk behaviour of large number of gas molecules.

In the thermodynamic description, spacetime will exhibit standard thermodynamic properties like entropy maximization, equipartition, thermal fluctuations etc. But all these thermodynamical features arise because the local Rindler observers attribute a density matrix to a pure quantum state after integrating out the unobservable modes. From this point of view, all these thermal effects are intrinsically quantum mechanical --- which is somewhat different from the `normal' thermal behaviour.  But our results suggest that this distinction between quantum fluctuations and thermal fluctuations is artificial (like e.g., the distinction between energy and momentum of a particle) and should fade away in the correct description of spacetime, when one properly takes into account the fresh observer dependence induced by the existence of local Rindler horizons.

Pursuing this idea further, and assuming that gravity is an intrinsically quantum mechanical phenomenon related to the microstructure of spacetime, one could try to relate the mechanical properties of gross matter to thermodynamics of spacetime. 
For example,
the mode function $\langle 0| \phi(x)|1_{\bf k}\rangle$ corresponding to
a one-particle state in either inertial frame or Rindler frame has a nonrelativistic ($c\to \infty$), quantum mechanical, limit in which the corresponding wave function 
 \begin{equation}
\psi(x) \propto \lim_{c\to \infty} e^{imc^2 t/\hbar} \langle 0|\phi(x) | 1_{\bf k}\rangle
\end{equation} 
 satisfies a Schrodinger equation with an accelerating potential $V=mgx$ when viewed from the second frame (see \cite{mach}). In describing the motion of a wave packet corresponding to such a particle,
  the quantum mechanical averages will satisfy the relation $\langle \delta E \rangle = mg \langle \delta x\rangle = F \langle \delta x\rangle$. On the other hand, given the thermal description in the local Rindler frame,
  we would expect a relation like $\langle \delta E \rangle = T\Delta S$ to hold suggesting that the entropy gradient $\Delta S$ (due to the gradient  $\Delta n$ in the microscopic degrees of freedom) present over a region $\langle \delta x\rangle$  to give rise to a force $F = T \Delta S/\langle \delta x\rangle$. If one assumes that (i) $\Delta S/k_B$ has to be quantized (based on the results of ref.\cite{entropyquant}) in units of $2\pi$ and (ii) $\langle \delta x\rangle \approx \hbar/mc$ for a particle of mass $m$, then we reproduce $F=mg$ on using the Rindler temperature $k_BT = \hbar g/2\pi c$. Alternatively, if one assumes that the force $F = T \Delta S/\langle \delta x\rangle$
should be equal to $mg$, then the universality of the Rindler temperature for bodies with different $m$ arises if we use $\langle \delta x\rangle = \hbar / mc$. In this case --- which involves  
 the quantum mechanical limit of a one-particle state in a non-inertial frame ---
 we need to handle simultaneously both quantum and thermal fluctuations. The expression $F = T \Delta S/\langle \delta x\rangle$  
 demands
  an intriguing interplay between thermal fluctuations (in the numerator, $T\Delta S$, arising from the non-zero temperature and entropy in local Rindler frame) and the quantum fluctuations (in the denominator, $\langle \delta x\rangle$, related to the intrinsic position uncertainty $\hbar/mc$) for the theory to be consistent, including the choice of numerical factors. (The idea that thermal and quantum fluctuations could be closely related to each other was earlier explored --- in a very different context --- in Ref. \cite{smolin}.)
  More generally, the observer dependence of entropy in non-inertial frames will require the development of  a unified formalism to handle the  thermal effects which arise due to the presence of horizons and the ``normal'' thermal phenomena of macroscopic bodies. 
  Such a development promises to pay rich dividends.

\end{document}